\pdfoutput=1
\documentclass[twocolumn,english,aps,superscriptaddress,prl]{revtex4-1}

\usepackage{babel}
\usepackage{amsmath}
\usepackage{amssymb}
\usepackage{graphicx}
\usepackage[colorlinks,bookmarks=false,citecolor=blue,linkcolor=red,urlcolor=blue]{hyperref}

\begin{document}

\title{Realization of higher Wess-Zumino-Witten models in spin chains}

\author{Fr\'ed\'eric Michaud}
\affiliation{Institute of Theoretical Physics, Ecole Polytechnique F\'ed\'erale de Lausanne (EPFL), CH-1015 Lausanne, Switzerland}
\author{Salvatore R. Manmana}
\affiliation{Institute for Theoretical Physics, University of G\"ottingen, D-37077 G\"ottingen, Germany}
\author{Fr\'ed\'eric Mila}
\affiliation{Institute of Theoretical Physics, Ecole Polytechnique F\'ed\'erale de Lausanne (EPFL), CH-1015 Lausanne, Switzerland}

\date{\today}
\begin{abstract} Building on the generalization of the exactly dimerized Majumdar-Ghosh ground state to
arbitrary spin $S$ for the Heisenberg chain with a three-site term $({\bf S}_{i-1}\cdot{\bf S}_{i})({\bf S}_{i}\cdot {\bf S}_{i+1})+H.c.$, we use density-matrix renormalization
group simulations and exact diagonalizations to determine the nature of the dimerization transition
for $S=1$, $3/2$ and $2$. The resulting central charge and critical exponent are in good agreement with the $SU(2)_{k=2S}$ Wess-Zumino-Witten values $c=3k/(2+k)$ and $\eta=3/(2+k)$. Since the 3-site term that induces dimerization appears naturally if exchange interactions are calculated beyond second order, these results suggest that $SU(2)_{k>1}$  Wess-Zumino-Witten models might finally be realized in actual spin chains.
\end{abstract}
\maketitle
\section{Introduction}
The solution in 1931 of the one-dimensional spin-1/2 Heisenberg model by Bethe\cite{bethe1931} is the first
example of a long series of exact solutions of integrable models that includes models of direct physical
relevance such as the one-dimensional Hubbard model\cite{lieb1967} and the Kondo problem\cite{andrei,wiegmann}.
As far as spin chains are concerned, there exist two families of integrable spin-S models which can be viewed
as generalizations of Bethe's solution. In both cases, the coupling is limited to nearest neighbours, and the coupling between spins at sites $i$ and $i+1$ takes the form of a polynomial of degree 2S in ${\bf S}_i\cdot{\bf  S}_{i+1}$. In the first family, the couplings are such that the Hamiltonian is the sum of permutations between nearest neighbours,
so that the symmetry is enlarged to $SU(N)$ with $N=2S+1$\cite{sutherland1975}. For this family, the central charge is equal
to $2S$, and the low-energy theory is equivalent to $2S$ free bosons. Another family of integrable models has been discovered in the early eighties in which the ground state is also critical\cite{takhtajan,babujian}, but the low energy field theory is the $SU(2)_{k=2S}$ Wess-Zumino-Witten model \cite{affleck_haldane}, with central charge $c=3k/(2+k)$\cite{alcaraz}. 
These two classes of model have been later on shown to belong to a single, more general family of integrable $SU(N)$ models
with higher symmetric representations\cite{andrei2,johannesson}, with
a low-energy sector described by the $SU(N)_{k}$ Wess-Zumino-Witten model, where $k$ is the rank of the symmetric irreducible
representation of $SU(N)$, as shown with Bethe ansatz\cite{martins} and confirmed numerically\cite{fuhringer}.
In both classes,
the Hamiltonian contains higher powers of ${\bf S}_i\cdot{\bf S}_{i+1}$ with significant coefficients, and the possibility
to realize them in actual spin chains is quite remote. For the first family of models, an alternative to spin chains is provided by ultracold fermionic alcaline earth atoms \cite{alcaline}, for which the $SU(N)$
Heisenberg model is the relevant effective model in the Mott insulating phase, but for the second family, there is no such alternative.

Shortly after the discovery of the second family of integrable models described by the $SU(2)_{k=2S}$ Wess-Zumino-Witten universality class,
it has been suggested that the simple spin-3/2 Heisenberg chain, which is expected to be critical,
might fall into this
category\cite{affleck1}, but this proposal has been definitely excluded by numerical simulations\cite{ziman}, which have shown that the
model has a central charge equal to 1. Since then, other models whose low-energy physics is described by the
$SU(2)_{k=2S}$ Wess-Zumino-Witten universality class have been discovered, including a generalization
of the Haldane-Shastry model to arbitrary spins\cite{greiter}, but all these models contain spin-spin interactions that seem
impossible to realize in actual systems, and the search for spin chain models that are both realistic and described
at low energy by the $SU(2)_{k=2S}$ Wess-Zumino-Witten universality class is still open.

Another route to this universality class, not followed by any concrete implementation so far, has been outlined by Affleck and Haldane\cite{affleck_haldane}, who argued on the basis of renormalization group arguments that the transition from a
uniform to a dimerized phase might take place through a critical point of universality class $SU(2)_{k=1}$ or $SU(2)_{k=2S}$ depending on the initial conditions of the flow. If the model is explicitly dimerized, the transition
is expected to be in the usual $SU(2)_{k=1}$ class, but if there is no explicit dimerization, as in the bilinear-biquadratic spin-1 chain, then the critical point might be $SU(2)_{k=2S}$.

In that respect, the recent generalization to arbitrary spin $S$ of the Majumdar-Ghosh exactly dimerized ground state of the $J_1-J_2$ spin-1/2 chain, where $J_1$ and $J_2$ are the exchange constants between nearest and next-nearest neighbors\cite{MG1,MG2}, opens new perspectives. Indeed, it has been shown that the spin-$S$ Heisenberg Hamiltonian with a three-site term
\begin{equation}
\label{hamiltonian}
\mathcal{H}  = \sum_i (J_1 {\bf S}_{i}\cdot{\bf S}_{i+1}+ J_{3}\left[\left({\bf S}_{i}\cdot{\bf S}_{i+1}\right)\left({\bf S}_{i+1}\cdot{\bf S}_{i+2}\right)+H.c.\right])
\end{equation}
is equivalent, for $S=1/2$, to the $J_1-J_2$ model with $J_2=J_3/2$, and that, for arbitrary $S$, the ground state
is exactly dimerized when $J_3/J_1=1/(4S(S+1)-2)$. Since the ground state of the pure Heisenberg
model ($J_3=0$) is not dimerized, there has to be a dimerization transition upon increasing $J_3$, and, since the model is not explicitly dimerized, the transition may have the more exotic $SU(2)_{k=2S}$ class for arbitrary $S$,
as already shown for $S=1$ \cite{MVMM}.

In this paper, using extensive density matrix renormalization group (DMRG) \cite{white,Schollwoeck1,Schollwoeck2} simulations and exact diagonalizations (ED), we show that this is indeed the case for $S=3/2$ and $S=2$. First of all, using a combination of level spectroscopy \cite{okamoto,okamoto2,schulz,gepner} and DMRG results for the size dependence of dimerization, we precisely locate the transition to dimerization,
with the conclusion that it occurs for quite small ratios of the coupling $J_3/J_1$ (0.063 and 0.0403 respectively). Then, using DMRG estimates of the central charge (extracted from the entanglement entropy) and of the spin-spin correlations, we show that, at the critical point, and
to a very good accuracy, the central charge is equal to $c=3 k/(2+k)$ and the exponent of the spin-spin correlation function is equal to $\eta=3/(2+k)$, with $k=2S$, in agreement with the $SU(2)_{k=2S}$ WZW universality class. So the
model of Eq. (\ref{hamiltonian}) appears to be a realistic one to implement the $SU(2)_{k=2S}$ WZW universality class in
actual spin chains.

{\it Numerical methods.---}
The main results of this paper have been obtained using DMRG simulations. 
To compute the correlation functions and the dimerization, we have used open boundary conditions. 
We have performed 12 sweeps keeping up to 1600 states in the last sweep. This has allowed us to reach good convergence up to 150 sites, with a discarded weight of the order of $10^{-8}$ close to the transition (and smaller far from the transition).

The calculation of the central charge is more difficult because it requires calculations with periodic boundary conditions, where DMRG is less efficient. We have deduced the central charge from calculations of the entanglement entropy on systems with $L=30,\ 40,\ 50$ and $60$ sites for $S=3/2$, and $L=30,\ 40$ and $\ 50$ sites for $S=2$. We did at least 7 sweeps, keeping 3500 states in the final sweep. Even with this high number of states, the maximum discarded weight was about $10^{-6}$. To get accurate estimates, we had to perform an extrapolation of the central charge with respect to the discarded weight, as explained below.

Exact diagonalizations of finite clusters have also been used in the context of the so-called level spectroscopy method\cite{okamoto,okamoto2,schulz,gepner} to locate the transition. To that end, we have used the Lanczos algorithm to compute the ground state energy in different sectors of total magnetization $S^z_{tot}$, momentum $k$ and parity under the transformation $i \rightarrow L-i$, where $i$ is the site label and $L$ the total number of sites. For spin $S=3/2$ chain, we have diagonalized systems with up to $L=16$, and for spin $S=2$ chains, we went up to $L=12$.

{\it Determination of the critical point.---}
We have used two different techniques to get an accurate estimate of the location of the transition between the uniform and the dimerized phase, one based on a direct calculation of the order parameter associated to the dimerization with DMRG, the other one based on exact diagonalizations.

\begin{figure}[t]
\includegraphics[width=0.5\textwidth]{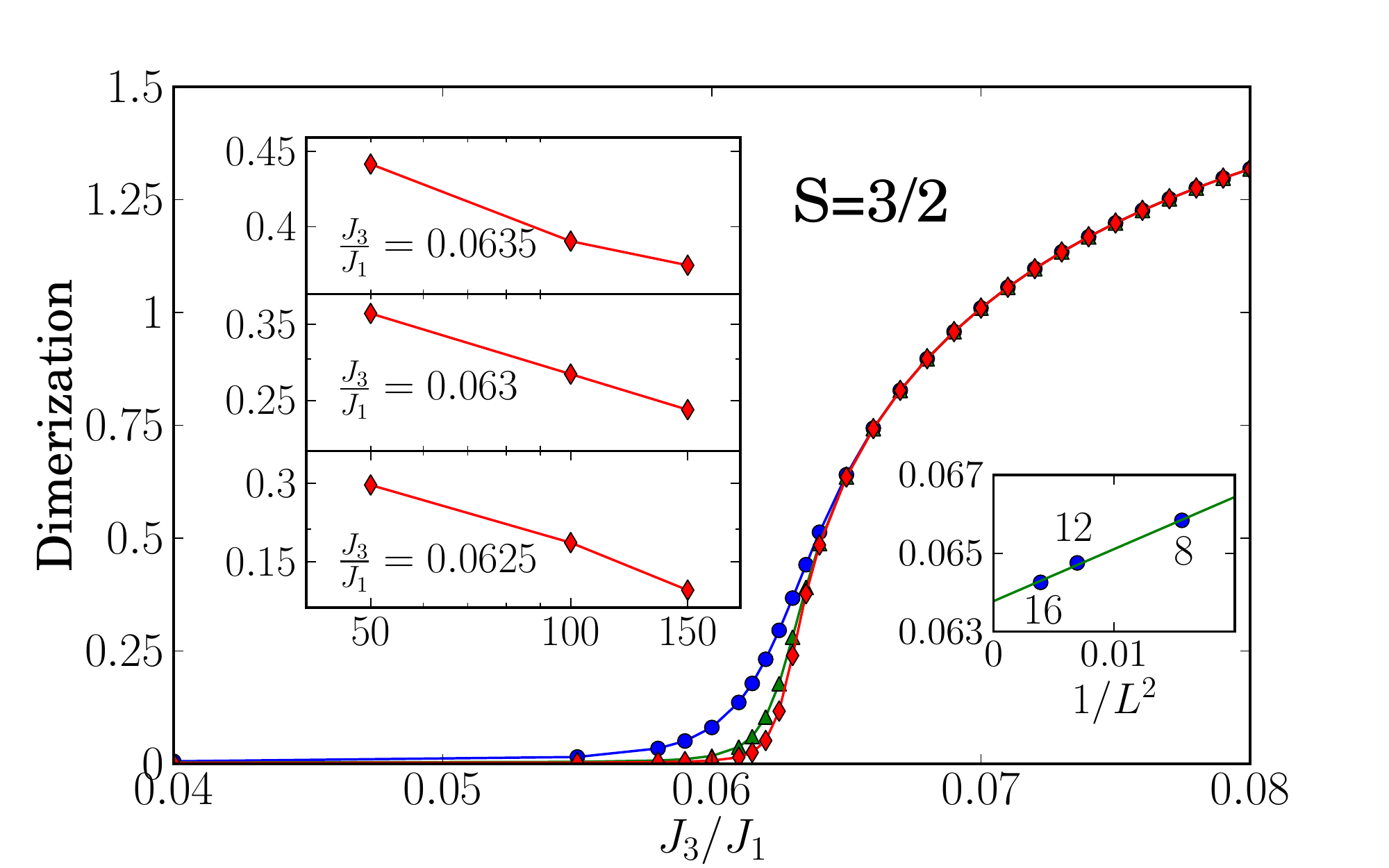}
\caption{(Color online) Dimerization as a function of $J_3/J_1$ for $S=3/2$ for different system sizes with up to $L=150$ sites in the vicinity of the phase transition $J_3/J_1 \approx 0.063$. The left insets show the size dependence at $J_3/J_1 = 0.0625, \, 0.063$ and $0.0635$, respectively. The right inset shows the localisation of the first excited state crossing point as a function of the system size.}
\label{fig:dim_S32}
\end{figure}

\begin{figure}[t]
\includegraphics[width=0.5\textwidth]{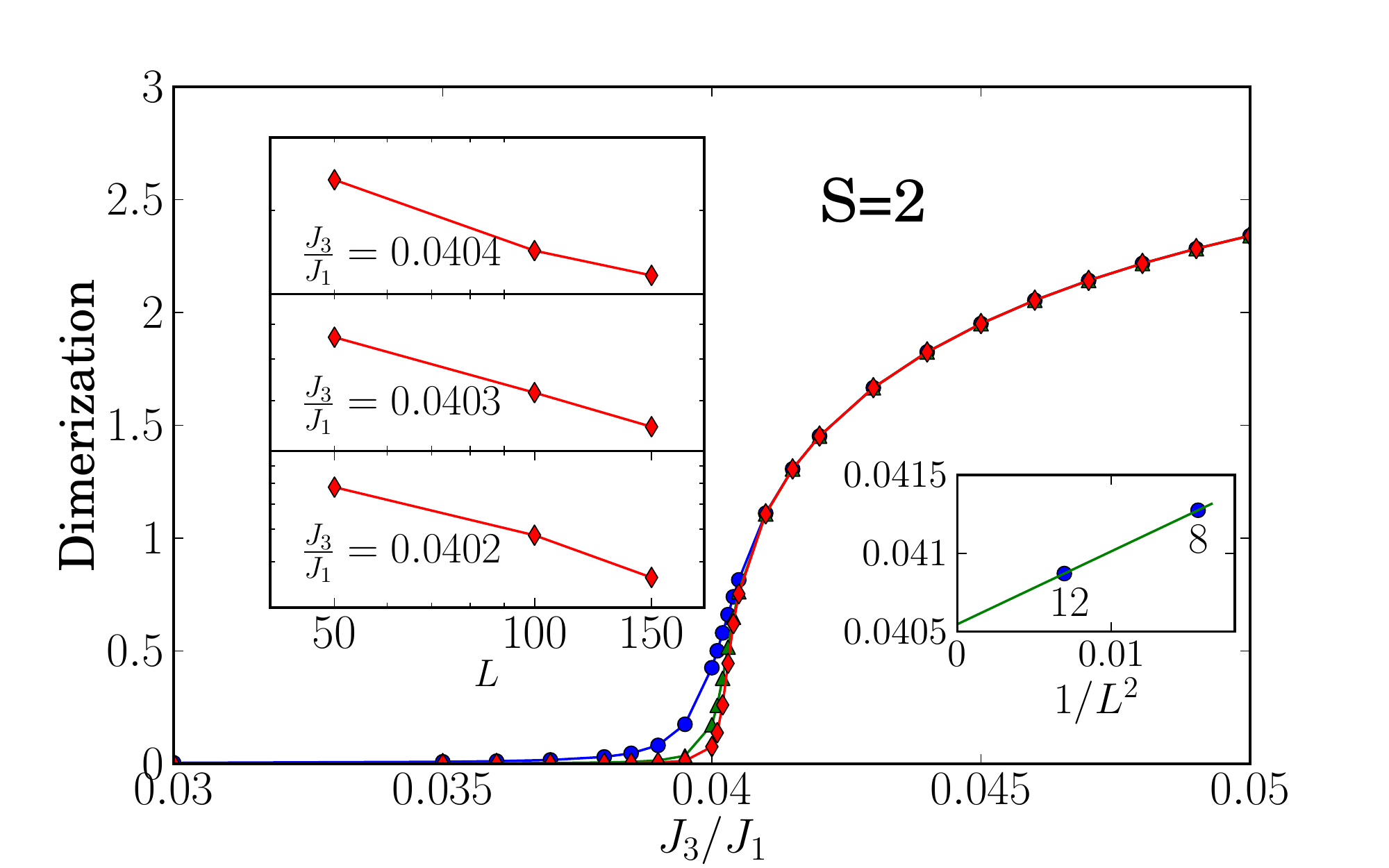}
\caption{(Color online) Dimerization as a function of $J_3/J_1$ for $S=2$ for different system sizes with up to $L=150$ sites in the vicinity of the phase transition $J_3/J_1 \approx 0.0403$. The left insets show the size dependence at $J_3/J_1 = 0.0402, \, 0.0403$ and $0.0404$, respectively. The right inset shows the localisation of the first excited state crossing point as a function of the system size.}
\label{fig:dim_S2}
\end{figure}

To determine the location of the phase transition with DMRG, we have computed the order parameter associated to the dimerization $d=\vert \vec S_{i-1} \cdot \vec S_i - \vec S_i \cdot \vec S_{i+1}\vert$. The order parameter in the center of the chain for different system sizes is shown in Fig. \ref{fig:dim_S32} for spin $S=3/2$ system, and in Fig. \ref{fig:dim_S2} for spin $S=2$. First of all, the data on finite-size systems
are in both cases consistent with a continuous transition. As for the spin-1/2 $J_1-J_2$ chain, the fact that the spin-3/2 chain is gapless below the transition, hence that the spin stiffness jumps at the transition, does not affect the continuous nature of the dimerization transition. In the thermodynamic limit, the dimerization order parameter is thus expected to be zero in the uniform phase, and to start growing at the transition with a critical exponent $\beta$ : $d \propto \left((J_3-{J_3}_c)/J_3\right)^{\beta}$. However,
it is not possible to locate the transition precisely using this scaling property because of finite-size effects. It is much more efficient to rely on the fact that, at the transition, the order parameter should tend to zero as a power law as a function of the system size, while it should tend to a finite value above it. Therefore, to obtain the transition point, we have plotted $d$ as a function of the system size $L$ on a log-log scale. At the transition, we expect a straight line, while above the transition we expect a convex behaviour.
This analysis is done in the left inset of Fig. \ref{fig:dim_S32} for spin $S=3/2$ and of Fig. \ref{fig:dim_S2} for spin $S=2$. It leads to a rather precise determination of the critical points at ${J_3}_c/J_1 = 0.063$ for $S=3/2$ and at ${J_3}_c/J_1 = 0.0403$ for $S=2$. Note that below the transition, the behaviour is expected to depend on the value of the spin: for integer spins, the spectrum is gapped, and the dimerization should decay exponentially, while for half-integer spin, the system is gapless with a Luttinger liquid exponent $K=1$, and the dimerization should decay as the inverse of the square of the size. Interestingly, the behaviours below the transition look very similar for spin 3/2 and 2, showing the same kind
of concavity. For spin 3/2, this is probably due to the fact that the fast decay only appears for larger sizes, so that for
the sizes available one only observes a cross-over regime, while for spin 2, the concavity reflects the exponential decay.
  
To check the reliability of these results, we have performed, following Ref.~\onlinecite{okamoto,okamoto2,schulz,gepner}, a level spectroscopy analysis. Since the transition is second order, we do not expect any level crossing in the ground state. However, the nature of the first excited state is expected to change at the transition \cite{okamoto}. For a system with $L=4n$, in the uniform region, the first excited state is a {\it triplet} with wave vector $\pi$ and parity $P=-1$, while in the dimerized region, the first excited state is a {\it singlet} with again wave vector $\pi$ but $P=1$. This method has been shown to lead to very accurate results \cite{okamoto} even if only rather small sizes are available 
because the finite-size scaling of the crossing point is very fast: $\alpha_c(L) = \alpha_c + \text{constant}/L^2+O(1/L^3)$, i.e. deviations scale as the {\it square} of the inverse size {\it without logarithmic corrections}. 

For spin $S=3/2$, we have calculated the first excited states for three system sizes ($L=8$, $12$ and $16$). The results are plotted in the right inset of Fig. \ref{fig:dim_S32}. The scaling with $1/L^2$ is quite accurate, and the level crossing extrapolates to ${J_3}_c/J_1 =  0.064$, in good agreement with the DMRG result ${J_3}_c/J_1 =  0.063$.
For spin $S=2$, we have calculated the first excited states for two system sizes ($L=8$, $L=12$). The results are shown in the right inset of Fig. \ref{fig:dim_S2}. The level crossing extrapolates to ${J_3}_c/J_1 =  0.0406$. The $1/L^2$ behavior could not be checked in this case because we have only two points. The result is nonetheless in good agreement with the DMRG result ${J_3}_c/J_1 =  0.0403$.

{\it Central charge.---}
At the transition between the uniform and the dimerized phase, it has already been shown that the $S=1$ chain with Heisenberg and three-body interactions can be effectively described by a $SU(2)_{k=2S}$ WZW model\cite{MVMM}. To check if this result can be extended to larger $S$, we have calculated numerically two quantities which are fixed by the $SU(2)_{k=2S}$ WZW universality class, namely the central charge, which is given by $c=3 k/(2+k)$, and the critical exponent of the decay of the spin-spin correlation function, which is given by $\eta=3/(2+k)$.

\begin{figure}[t]
\includegraphics[width=0.5\textwidth]{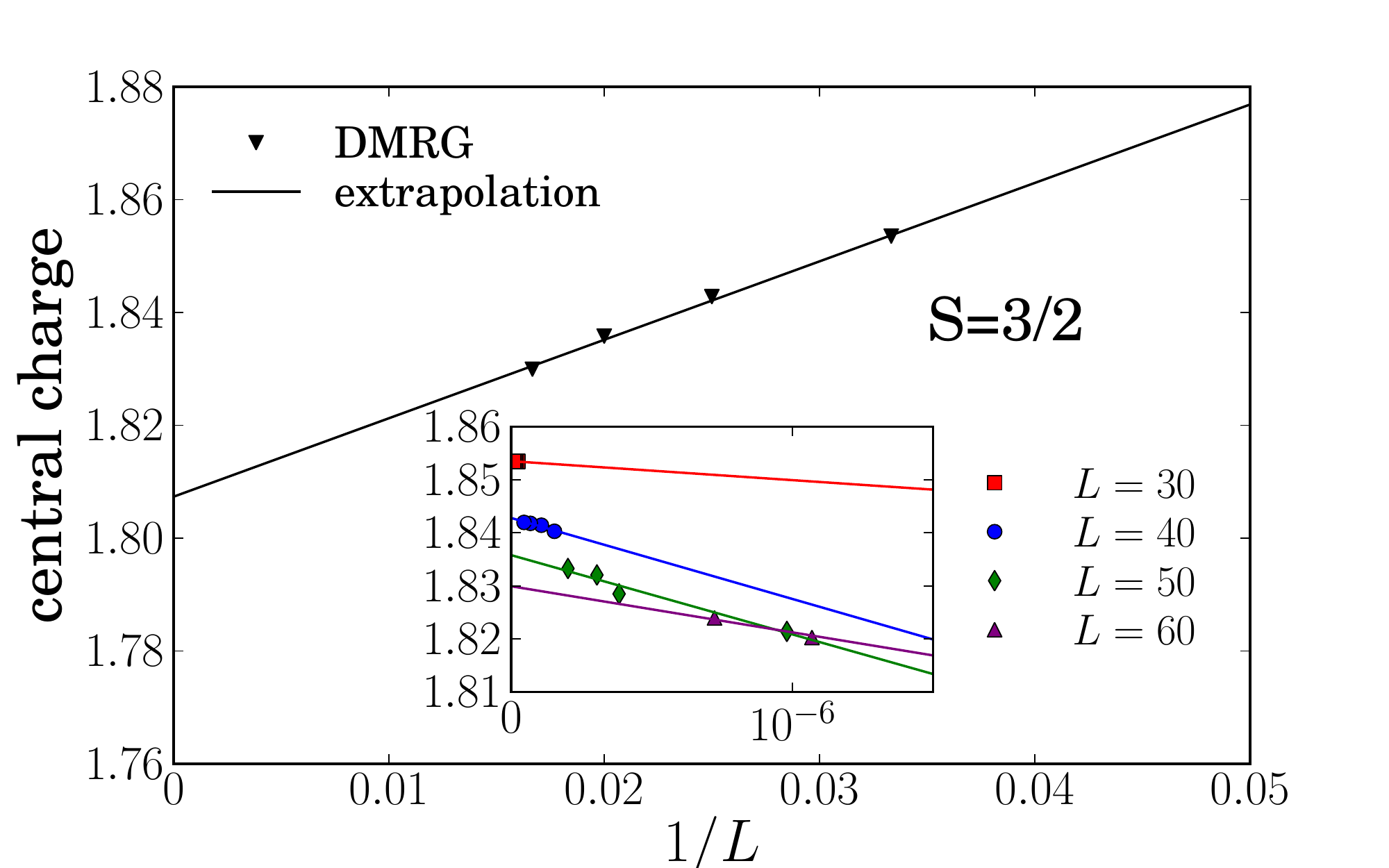}
\caption{(Color online) Extrapolation to the thermodynamic limit of the central charge for $S=3/2$. The data points shown are results from an extrapolation as a function of the discarded weight, which is displayed in the inset.}
\label{fig:cc_S32}
\end{figure}

\begin{figure}[t]
\includegraphics[width=0.5\textwidth]{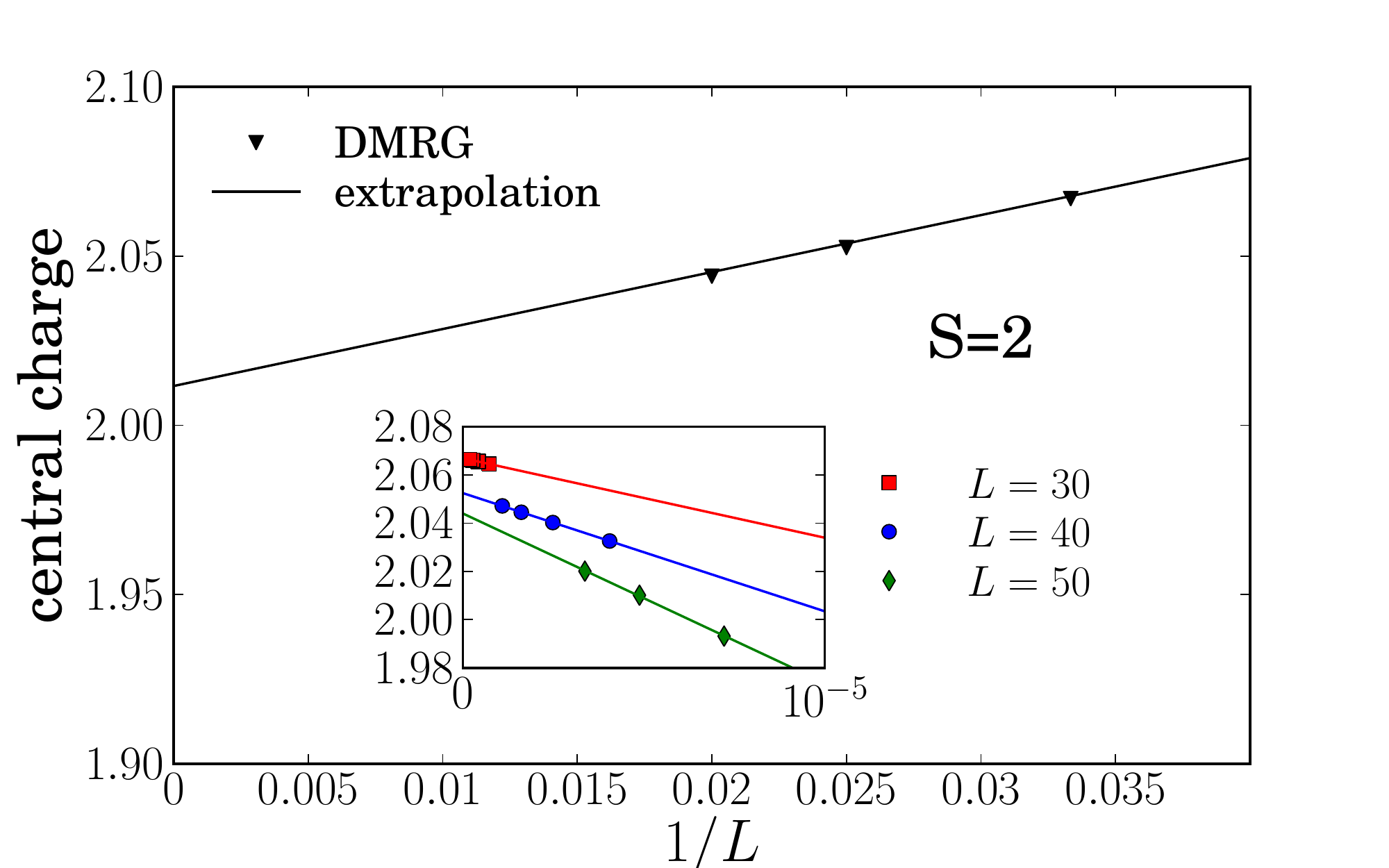}
\caption{(Color online) The same as in Fig. \ref{fig:cc_S32} for S=2.}
\label{fig:cc_S2}
\end{figure}

The central charge can be extracted for a finite system from the entanglement entropy obtained with DMRG on a system with PBC. More precisely, it is obtained by fitting the numerical results with the following formula \cite{cardy}:
\begin{equation}
S_\ell = \frac{c}{3} \ln \left[ \frac{L}{\pi} \sin \left( \frac{\pi \ell}{L}\right) \right] + g_{\rm PBC},
\label{eq:entropy}
\end{equation}
where  $c$ is the central charge, $S_\ell = - {\rm Tr} \varrho_\ell \ln \varrho_\ell$ the entanglement entropy, and $\varrho_\ell$ the reduced density matrix of a subsystem of size $\ell$.
This analysis should then be followed by a finite size scaling of $c$. However, due to the rapid increase of the {\it cpu}
time required, we have not been able to reach full convergence at the critical point for $30$ sites and beyond. 
For a given
size, we have therefore kept track of the discarded weight at various steps of the simulation, and we have performed an extrapolation of the central charge with respect to the discarded weight. A linear function turned out to fit well the results as shown in the inset of Fig \ref{fig:cc_S32} for $S=3/2$ and in the inset of Fig. \ref{fig:cc_S2} for $S=2$, leading to estimates of the central charge for different sizes. This analysis has been performed for $L=30,40,50,60$ for $S=3/2$ and $L=30,40,50$ for $S=2$, and a finite-size scaling of the result has revealed a linear
behaviour as a function of $1/L$. For spin $S=3/2$, the extrapolation to the thermodynamic limit is shown in Fig. \ref{fig:cc_S32}. The $SU(2)_{k=3}$ WZW theory predicts a central charge of $9/5=1.8$. Our numerical extrapolation leads to $c=1.807$, which is remarkably close to the expected value, leaving little doubt about the nature of the transition. For $S=2$, the extrapolation shown in Fig. \ref{fig:cc_S2} leads to $c=2.01$, again in very good agreement with the value $c=2$ for the $SU(2)_{k=4}$ WZW model. These results nicely confirm that the transition is also in the  $SU(2)_{k=2S}$ WZW universality class for S=3/2 and S=2.

\begin{figure}[t]
\includegraphics[width=0.5\textwidth]{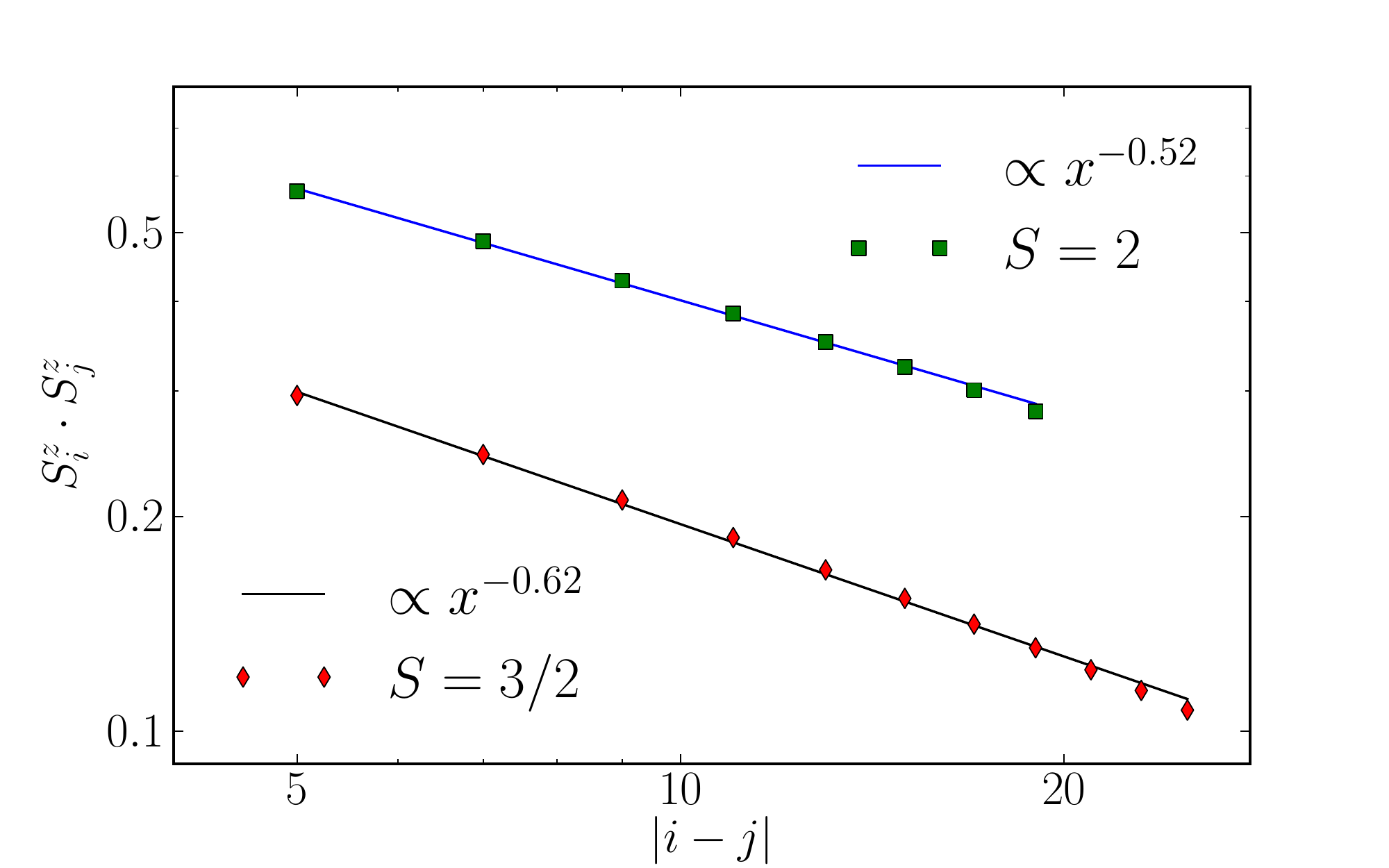}
\caption{(Color online) Decay of the spin correlation function $\langle S_i^z \cdot S_j^z \rangle$ with distance $|i-j|$ for systems with $L=150$ and spin $S=3/2$ and $S=2$. The solid lines are fits to power laws with exponents indicated in the plot. }
\label{fig:corr}
\end{figure}

{\it Critical exponent of the spin-spin correlation function.---}
We now turn to the extraction of the critical exponent associated to the spin-spin correlation function defined by $C_s(i,j)=\langle S^z_i \cdot S^z_j \rangle$. At the transition, it is expected to decay as a power law according to $ C_s(i,j) \propto \left (   i - j  \right)^\eta$. In such a case, the boundary effects are expected to play an important role, and one can only hope to observe the bulk behavior not too close to the boundaries. In practice, we could only perform a fit to determine $\eta$ inside a narrow window in the centre of the chain where the decay looked polynomial. In the fit, we only considered odd distances to get rid of the inaccuracy coming from the odd-even oscillation in the correlation function.The extrapolations are shown in Fig. \ref{fig:corr}. For $S=3/2$, we have obtained $\eta=0.62$. This has to be compared to the WZW model where $\eta=3/5$. So, within a small region, the correlation function seems to be well described by the WZW model. The same analysis for $S=2$ leads to an exponent $\eta=0.52$, which should be compared to the result $\eta=1/2$ of the WZW theory. These results are therefore compatible with a $SU(2)_{k=2S}$ WZW theory.

{\it Discussion and conclusion.---}
To summarize, we have shown that, for $S=3/2$ and $S=2$, the dimerization transition in the $J_1-J_3$ model between a uniform and a spontaneously dimerized phase is in the $SU(2)_{k=2S}$ WZW universality class. The evidence relies on a careful extrapolation of the central charge at the transition, which agrees with the theoretical prediction $c=3 k/(2+k)$ with an error smaller than 1 per cent (see Table \ref{table:summary}). To further confirm this conclusion, we have computed the exponent $\eta$ of the spin-spin correlation function, and the results are again compatible with a WZW critical point. All the numerical results obtained in this paper, and in a former paper \cite{MVMM} for $S=1$, are summarized in Table \ref{table:summary}.

In view of these results, it is tempting to conjecture that the universality class of the dimerization transition will be $SU(2)_{k=2S}$ 
for all values of $S$. This is likely to be the case if the transition is continuous. However, as pointed out in the context of one-dimensional coupled dipolar gases, the fixed point might be unstable for $S>2$ because of the presence of relevant operators that are forbidden by symmetry for $S\le 2$\cite{lecheminant}, leading to a first order transition, in agreement with recent numerical results\cite{tsvelik_kuklov}. This issue, which requires to investigate spins
larger than 2, is left for future investigation.

\begin{table}[h!b!p!]

\begin{tabular}{|l|l|l|l|l|}
\hline
 &  $S=1$ &  $S=3/2$ & $S=2$ & $2 S = k$ \\
\hline
${J_3}_c/J_1$  (DMRG) &  $0.111$ &  $0.063$ & $0.0403$ & $\textendash $\\
\hline
${J_3}_c/J_1$   (ED) &  $\textendash$ &  $0.064$ & $0.0406$ &$\textendash$ \\
\hline
$J_3/J_1$ (MG point) &  $1/6$ &  $1/13$ & $1/22$  & $1/(4S(S+1)-2)$\\
\hline
$c$ (WZW) & $3/2$ &  $9/5$ & $2$ & $3 k/(2+k)$ \\
\hline
$c$ (DMRG) &  $1.502$ &  $1.807$ & $2.01$& $\textendash$ \\
\hline
$\eta$ (WZW) &  $3/4$ &  $3/5$ & $1/2$ & $3/(2+k)$ \\
\hline
$\eta$ (DMRG) &  $0.72$ &  $0.62$ & $0.52$ & $\textendash$ \\
\hline
\end{tabular}
\caption{\label{table:summary}Summary of all quantities computed with DMRG and the same quantities in the WZW model: the ratio $J_3/J_1$
at the dimerization transition and at the MG point, the central charge $c$ and the exponent $\eta$ of the 
spin-spin correlation function. The data for S=1 are from Ref.~[\onlinecite{MVMM}].}
\end{table}

Finally, by contrast to the spin $S=1/2$ $J_1-J_2$ chain where the critical point is the standard $SU(2)$ level 1 field theory, the physics is not standard for higher spin. This is particularly clear for spin $S=3/2$, where the central charge takes the rather exotic value $c=9/5$.  To our knowledge, up to now no simple and realistic spin chain model which allows to probe this kind of physics has been put forward. In that respect, the Hamiltonian we propose has many advantages: it is simple - it has only two different terms and interactions do not reach beyond next-nearest neighbors - and it is realistic since the three-site term emerges naturally from a strong coupling expansion of a multi-orbital Hubbard model. It is thus our hope that the results of the present paper
will prove useful on the way to an experimental realization of higher WZW models in spin chains.

We acknowledge useful discussions with D. Cabra, F. Essler and T. Ziman. We are especially indebted to P. Lecheminant for several interesting comments, in particular for pointing out to us the possibility of a first order transition for spins larger than 2. 
This work has been supported by the Swiss National Fund and by MaNEP.

\bibliographystyle{prsty}
\bibliography{bibliography}

\end{document}